# ASSESSMENTS OF ALI, DOME A, AND SUMMIT CAMP FOR MM-WAVE OBSERVATIONS USING MERRA-2 REANALYSIS

CHAO-LIN KUO

Physics Department, Stanford University, 385 Via Pueblo Mall, Stanford, CA 94305, USA;
clkuo@stanford.edu

SLAC National Accelerator Laboratory, 2575 Sand Hill Rd, Menlo Park, CA 94025, USA

## ABSTRACT

NASA's latest MERRA-2 reanalysis of the modern satellite measurements provides unprecedented uniformity and fidelity for the atmospheric data. In this paper, these data are used to evaluate five sites for millimeter-wave (mm-wave) observations. These include two established sites (South Pole and Chajnantor, Atacama), and three new sites (Ali, Tibet; Dome A, Antarctica; and Summit Camp, Greenland). Atmospheric properties including precipitable water vapor ($PWV$), sky brightness temperature fluctuations, ice and liquid water paths are derived and compared. Dome A emerges to be the best among those evaluated, with $PWV$ and fluctuations smaller than the second-best site, South Pole, by more than a factor of 2. It is found that the higher site in Ali (6,100 m) is on par with Cerro Chajnantor (5,612 m) in terms of transmission and stability. The lower site in Ali (5,250 m) planned for first stage of observations at 90/150GHz provides conditions comparable to those on the Chajnantor Plateau. These analyses confirm Ali to be an excellent mm-wave site on the Northern Hemisphere that will complement well-established sites on the Southern Hemisphere. It is also found in this analysis that the observing conditions at Summit Camp are comparable to Cerro Chajnantor. Although it is more affected by the presence of liquid water clouds.

*Subject headings*: atmospheric effects — cosmic microwave background — site testing

## 1. INTRODUCTION

Searching for imprints of primordial gravitational waves using CMB polarization [1] and probing the accretion properties of blackholes using very long-baseline interferometry (VLBI) [2] are two of the most exciting topics in modern astrophysics. These two endeavors share similar scientific themes in that CMB polarization probes the beginning of time, and VLBI such as the Event Horizon Telescope (EHT) studies the edge of space. Experimentally, scientists working in these two areas must make sensitive measurements through the atmosphere in the handful of millimeter-wave (mm-wave) frequency bands. In these atmospheric "windows", the opacity depends strongly on observing conditions, particularly the integrated humidity. The most notable mm-wave atmospheric windows (3mm, 2mm, 1.4mm, 1.1mm; or 90 GHz, 150 GHz, 220GHz, and 280 GHz) near the peak of the CMB radiation can be rendered opaque if the $PWV$ (precipitable water vapor) is larger than a few mm. Not only do water vapor and atmospheric column density introduce attenuation in the astronomical signal, Kirchhoff's law of thermal radiation demands that the emission from the intervening media, and therefore the thermal background against which the measurements are made, is proportional to the product of the optical depth and the physical temperature (>200K). This latter effect makes it desirable to observe at sites with the lowest possible opacity in the mm-wave windows, even when the attenuation of the signal starts to become a secondary factor when the transmission is larger than 90%.

In addition to adding a constant background, the concentration of water – the most important mm-wave absorber in the atmosphere – tends to fluctuate because the saturated water vapor pressure depends strongly on the temperature. As a result, the distribution of water in different phases greatly amplifies the density fluctuations in the dry air.

Vapor-induced fluctuations in optical depth create brightness temperature fluctuations in CMB measurements and introduce phase errors in long-baseline interferometry.

So far, ground-based CMB measurements have been made most successfully at the South Pole and the Atacama Desert. EHT has also gathered data from both sites. In addition to science data collected, the observing conditions at these sites have been characterized over the last few decades using a variety of techniques, including coherent radiometers with frequencies tuned to the wings of the water lines, far-infrared (IR) continuum photometers, and radiosonde launches that collect temperature, humidity, and pressure data twice a day.

Recently, the CMB and VLBI communities became interested in exploring other sites to seek either larger sky area or more complete *uv* coverage [3]. Science with CMB lensing science especially benefits from large sky coverage that includes a northern site [4]. Summit Camp in Greenland and Ali, Tibet (near the town of Shiquanhe) have been proposed as candidate sites in the Northern Hemisphere. Another site with great potential is Dome A, the highest point of Antarctic Plateau (4,093m) where the Chinese Kunlun Station is located. Recent FTS data showed that it is possibly the best THz site on Earth [5].

Efforts are being made to directly characterize the all three sites using various instruments. In this paper, the MERRA-2 reanalysis [6] produced by NASA's Global Modeling and Assimilation Office (GMAO) have been used to evaluate these sites. While onsite radiometer measurements are not to be replaced by the "armchair" approach described in this paper, the most recent MERRA-2 data products derived from IR through microwave provide unparalleled completeness and uniformity of the atmospheric properties, making it very straightforward to evaluate different sites on an equal footing. In this paper, MERRA-2 analysis products have been used to derive *PWV* measurements with intervals as short as 1 hour. These results are found to be in excellent agreement with prior measurements at South Pole and Atacama. The same analyses are carried out for Ali, Summit Camp, Dome A.

This paper outlines a recipe to use the MERRA-2 data products to comprehensively evaluate candidate mm/sub-mm sites. Dome A is confirmed to be the premier mm/submm sites on Earth, with exceedingly low *PWV* and brightness temperature fluctuations all year around. The proposed Northern Hemispheric sites Ali, Tibet and Summit Camp, Greenland are comparable to Atacama, although both sites are noticeably cloudier. Before this paper, 2D weather data from CFSR have been used to evaluate the Ali site [7]. Although the results in that work show initial promise of Ali, they are limited by a combination of grid resolution and the lack of altitude information. Data taken from a remotely operated 220 GHz tau-meter have been reported for the Summit Camp by the ASIAA (Institute of Astronomy and Astrophysics, Academia Sinica) group [8]. Dome A has recently been characterized with an FTS (Fourier Transform Spectrometer) installed by Purple Mountain Observatory (PMO) and Harvard-Smithsonian Center for Astrophysics (CfA) [5]. The results presented in this paper are in excellent agreement with these heroic efforts, but with more emphasis on long integrations such as CMB and other deep-field mm-wave measurements. To the author's knowledge, this is the first time the effects of clouds, quantified by *IWP* (ice water path) and *LWP* (liquid water path), are systematically assessed for mm-wave observation. Finally, this paper presents the first comprehensive comparisons of brightness temperature fluctuations between Dome A, South Pole, Atacama, Ali, and Summit Camp with a uniform methodology.

2. DATA AND ANALYSIS

The Modern-Era Retrospective Analysis for Research and Applications, Version 2 (MERRA-2) is the latest atmospheric reanalysis of the modern satellite era produced by NASA's Global Modeling and Assimilation Office (GMAO). It uses state-of-the-art processing of observations from the continually evolving global observing system that employs an array of instruments from near-IR to microwave [9]. The redundantly observed atmospheric quantities have been forced to obey the equation of continuity and other hydrodynamical relations using large scale numerical modeling. Data taken from different instruments, or retrieved with different methodologies, have been extensively compared and calibrated. MERRA-2 data are also made to be consistent with other atmospheric data such as radiosonde launches when they are available. Altogether, this dataset offers reliable measurements of basic meteorological fields such as temperature, pressure, specific humidity, ice/liquid cloud fraction, wind speed/direction in *three* dimensions, as well as other atmospheric quantities ($CO_2$, $O_3$, aerosol) that are more relevant to astronomers than to astronomy.

Several MERRA-2 products provide sufficient pressure resolution to map the low tropospheric profiles, easily resolving the Antarctic inversion layer in both temperature and specific humidity (Figure 1). Previous MERRA products lacks such resolution to reliably calculate *PWV* [5].

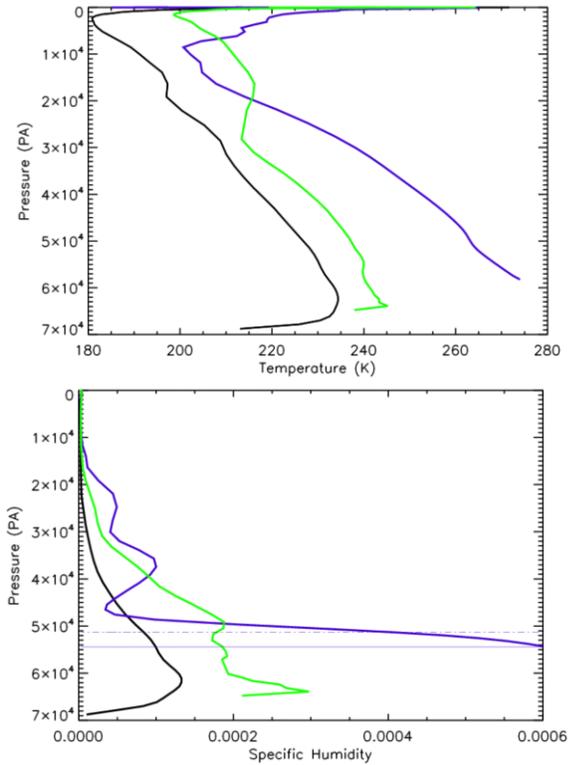

Figure 1. Example wintertime temperature (*Top*) and humidity (*Bottom*) profiles from MERRA-2 reanalysis. The three lines are for South Pole (black), Summit Camp, Greenland (green), and Chajnantor (blue). The strong Antarctic surface inversion is clearly resolved in both temperature and humidity. The horizontal lines mark the approximate elevation of Simons Observatory and Cerro Chajnantor.

As mentioned earlier, *PWV* distribution is the most important measure for site quality. But MERRA-2 offers a lot more information that are important for mm-wave observations. For example, the presence of ice and liquid clouds can impact CMB observations that rely on relentless integrations with high observing duty cycles, but is often overlooked in conventional site surveys focusing on opportunistic VLBI or THz spectroscopy of bright sources. MERRA-2 provides measurements of *IWP* and *LWP* which, when measured in $kg/m^2$, are equivalent to the same molecular column density for mm of *PWV* in solid and liquid phases. Through the analyses of MERRA-2 data, the impact of clouds – particularly clouds consisting of supercooled liquid droplets – on CMB observations for these sites are reported.

MERRA-2 published many analysis products for different applications in atmospheric modeling and climate monitoring. All the results in this paper are derived from four data products: (1) 3D three-hour interval, time-averaged meteorological fields with 72 model pressure levels (M2T3NVAS M.5.12.4), *e.g.* Figure 1; (2) same as the above, but for instantaneous quantities (M2I3NVAS M.5.12.4); (3) 2D integrated diagnostic meteorological fields with 1-hour time intervals, time-averaged (M2T1NXSLV.5.12.4); (4) same as the above for instantaneous quantities (M2I1NXAS M.5.12.4). All four data products come with a grid resolution of 0.65°×0.5° in (longitude, latitude).

The data products with higher pressure resolution (1)(2) are chosen to capture the temperature and humidity near the surface, especially for the Antarctic sites. However, since the spatial resolution of MERRA-2 is insufficient to resolve individual peaks for siting observatories, the altitude data in (1) and (2) are used to evaluate potential sites at different elevations. This is done under the assumption (the "elevation hypothesis") that the variations within a MERRA-2 grid are mostly due to the elevation, at least on average. Specifically, the *PWV* on a mountaintop within the grid point has been calculated as the integral of specific humidity starting at the altitude (pressure) of the candidate site. As discussed in Section 4, *PWV* calculated using this method are consistent with prior data taken at Cerro Chajnantor and Simons Observatory, two sites located within the same grid point but at different elevations. The same assumptions are made in evaluating candidate sites within the Ali grid point, at 5,250 m (Ali1) and 6,100 m (Ali2). For South Pole, Dome A, and Summit Camp, the ground elevation does not vary significantly within a MERRA-2 grid point.

The PWV can be calculated as the integral of specific humidity $q$ as a function of pressure $p$:

$$PWV = \frac{1}{\rho g} \int_{p_0}^{0} q \, dp, \quad (1)$$

where $p_0$ is the pressure at the elevation.

Brightness temperature fluctuations introduce phase noise in VLBI and $1/f$ noise in polarization data in scanning/differencing CMB experiments (when relative gain error is present). While the spatial resolution of MERRA-2 (~50 km) is insufficient to directly infer sky temperature fluctuations on angular scales that are relevant to astronomical measurements, the richness of MERRA-2 data products offers several ways to estimate brightness temperature fluctuations for these purposes.

MERRA-2 3D data are provided at 3-hour intervals. Variance of inferred *PWV* provides information on the sky brightness variations (and phase stability) longer than this time scale. On the other hand, differencing the 3D instantaneous and time-averaged specific humidity ( $q_i$ and $q_a$ ) provides a measurement of instantaneous *spatial* variations with characteristic scales of a few hundred meters (Figure 3). Since these scales are still within the 3D regime of Kolmogorov-Taylor turbulence, same fluctuations should appear as the angular brightness temperature variations across the sky. This method is analogous to using redshift information to map the large-scale structure. Specifically, fluctuations on scales of 200 m appears as ~10° (multipole $l$~20) angular variations, which are very relevant to measurements of primordial *B*-modes. The amplitude of the fluctuations defined this way is given by

$$dPWV \equiv \frac{1}{\rho g}\left[\int_{p_0}^{0}(q_i - q_a)^2 p \, dp\right]^{\frac{1}{2}}. \quad (2)$$

The corresponding brightness temperature fluctuations at each frequency are calculated by the radiative transfer code *am* [10].

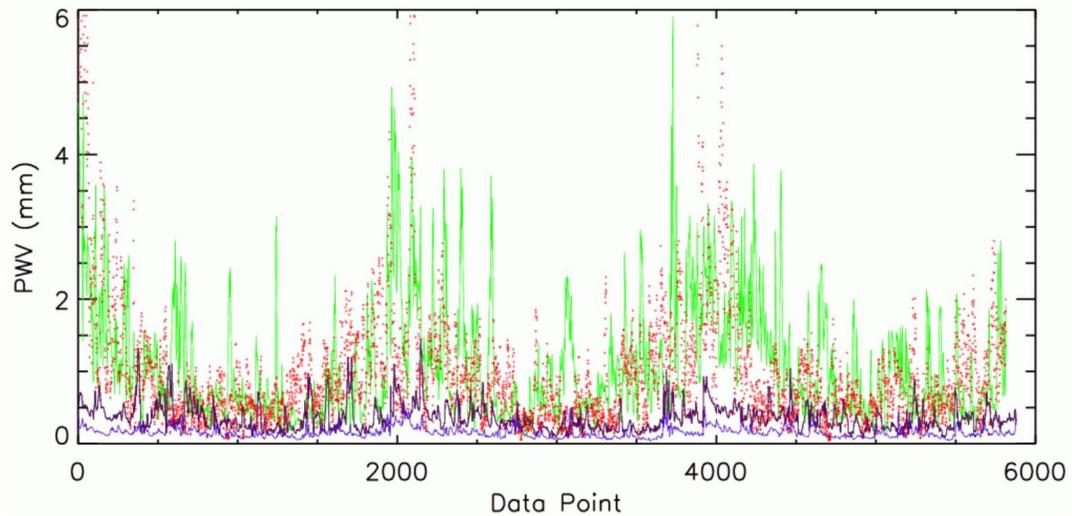

Figure 2. The calculated *PWV* for South Pole (black), Dome A (blue), Greenland (green), and Ali2 (red). The 4 summer months have been excluded for all these sites, but otherwise the data are continuous, 3-hr per point, without any gaps.

This paper also uses another method to reveal short-time-scale fluctuations. For the 2D data with 1-hour intervals, MERRA-2 supplies both the instantaneous and time-averaged data. Because the instantaneous data contains *aliased* noise but the time-averaged data do not, differencing these data provides measurements of temporal variations taking place shorter than two hours, *e.g.*, for *PWV* and *LWP* (denoted by $\Delta PWV, \Delta LWP$, respectively). This is especially useful for South Pole and Summit Camp, where the 2D data are directly applicable due to the flat terrain.

It should also be noted that temporal variations are closely related to angular brightness variation in the screen model of the atmospheric fluctuations, in which the large-scale wind carries a "screen" of spatially varying distribution of water vapor and droplets [11,12]. In the past, this model successfully describes the behavior of brightness temperature fluctuations and phase noise at mm-waves. For a wind speed of 2 m/s and screen height of 1 km, 30-minute-long temporal variations correspond to brightness fluctuations on the largest angular scales observable (>~90°), or multipole $l$ <5.

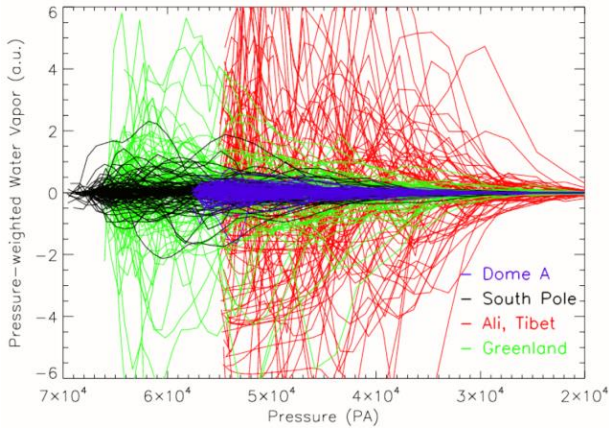

Figure 3. The spatial fluctuations of water vapor measured by MERRA-2. Fluctuations on scales of a few hundred meters seen in this plot are used to derived *dPWV* for site evaluation.

While $d$ and $\Delta$ quantities are not designed to precisely describe the angular power spectra, they should be good proxies for angular fluctuations of brightness temperature relevant for CMB observations. These analyses can be used to provide reliable comparison of brightness temperature fluctuations for different sites, which is the main purpose of this paper and other similar endeavors.

3. IMPACT ON OBSERVATIONS

Water is a powerful absorber in mm/submm wavelengths. While water vapor makes up a small fraction of the atmosphere, *PWV* is the most important quantity that characterize the quality of the site for mm-wave and THz observations. At frequencies higher than 150 GHz, observations are only effective when *PWV* drops below 1 mm. Since superconducting transition edge sensors (TES) have become standard CMB detectors, reliable measurements of *PWV* distribution became even more important, because TES can only operate under optical loading up to certain saturation power, above which the responsivity of TES drops rapidly. A full distribution (histograms) of *PWV* can be used to tune the thermal conductance ($G$) and critical temperature ($T_c$) to optimize the sensitivity of the experiment over extended period of integration. Without such information, conservatively parameters that degrade sensitivity are often chosen to avoid detector saturation.

Compared to water vapor, non-precipitating ice/liquid water (*i.e.,* clouds) makes up an even smaller mass fraction. Liquid droplets and ice crystals back-scatter thermal radiation from Earth's warm surface into the line-of-sight of the telescope, increasing the observed $T_b$. Because of their extended shapes, ice crystals scatter strongly in submm and far-IR. At mm-wave and microwave where Rayleigh approximation holds, liquid droplets scatter more strongly than ice crystals because of the higher index of refraction. For the same molecular column density, liquid water is much more emissive than vapor or ice. Therefore, even very thin liquid water clouds are very detrimental to mm-wave observations (Figure 4). Unfortunately, supercooled liquid clouds exist way below "freezing", down to a temperature of ~228K. Based on the MERRA-2 data, *LWP* often coexist with *IWP* and starts to impact mm-wave brightness temperature when *PWV* is greater than about 0.5mm, especially for Ali and Summit Camp. The effects of *IWP* (cirrus clouds) are found to be small for all the sites considered in this paper.

MERRA-2 data have been used in [10] to derive the mean temperature, humidity, and ozone profiles as the inputs to radiative transfer calculations (*am*) for different sites, including the ones considered here. In this paper, each set of meteorological measurements have been directly used for an *am* calculation. While this requires many more calls, the $T_b$ distribution obtained this way explicitly include all the correlations between pressure levels and across different quantities (such as temperature, humidity, and clouds).

For CMB, sky temperature fluctuations do *not* fundamentally impact the polarization measurements. If the fluctuations are unpolarized, they can be reduced by better gain-fitting, or by rapid modulations of polarization by a Half-Wave-Plate. But at present, brightness temperature fluctuations on time scales longer than a few seconds still impose a practical limitation on science grade CMB polarization measurements. Results from the fluctuation analysis presented in this paper should be very useful for future planning such as the CMB-S4 effort [13,14].

Atmospheric fluctuations in polarization can be generated by patches of ice crystals descending through air, or aligned by varying wind. Such polarization fluctuations are an irreducible source of noise, because they cannot be mitigated by either method mentioned above. While MERRA-2 data provide measurements of *IWP*, no attempts are made in this paper to estimate the degree of polarization of the scattered ground radiation, which depends on the shapes and sizes of the crystals and the degree of alignment. Such analysis is better done as an empirical cross-correlation between mm-wave polarization signal and *IWP*, and is beyond the scope of this paper.

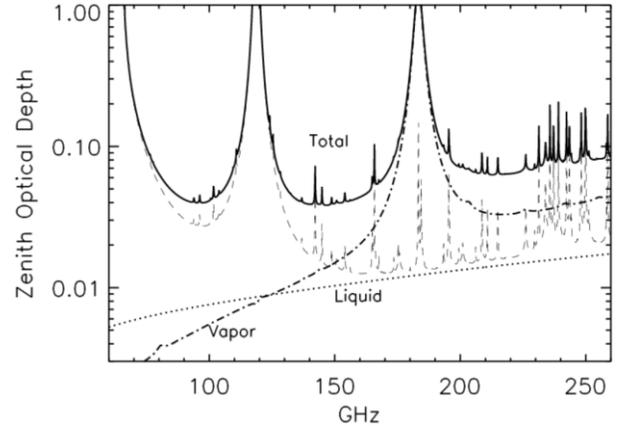

Figure 4. *am* spectrum for the dry and wet components of the atmosphere (*PWV*=0.65mm, *LWP*=0.01 kg/m$^2$).

Although liquid droplets are spherical, they can scatter an anisotropic radiation field (specifically, the quadrupole moment) into the line-of-sight with a small amount of polarization. This process is analogous to the generation of *E*-mode polarization at the surface of last scattering [15]. Since the large-scale radiation field experienced by water molecules ($T = T_{sky}$ for $\theta < \pi/2$; $T = T_{gnd}$ for $\theta > \pi/2$) lacks any quadrupole moments, the typical amplitude of the quadrupole moment is at the 10s of $mK$ level, which leads to negligible level of polarization.

Finally, it should be noted that dry air has a significant contribution to opacity at 90 GHz, a quintessential CMB channel, because of proximity to a strong oxygen line. Observations made in this window are particularly advantageous at higher mid-latitude sites (Ali and Atacama), somewhat compensating for their higher *PWV* compared to the Antarctic sites.

## 4. RESULTS

In this section, results from the MERRA-2 analyses, including the *PWV, IWP, LWP, $T_b$*, and their fluctuations are presented, first for Atacama and South Pole as a validation, and then for the proposed sites. The data are taken from three years of MERRA-2 reanalysis (2014–2016), excluding summer months (May–August for Northern Hemisphere and November–February for Southern Hemisphere).

## 4.1 MERRA-2 Properties of The Established Sites

*Atacama Desert*   The Atacama Cosmology Telescope (ACT) [16] and POLARBEAR Telescope [17] (jointly the Simons Observatory, SO) are located at an elevation of 5,190 m in Cerro Toco, about 200 m higher than the Chajnantor Plateau, where ALMA (Atacama Large Millimeter Array) is located. Yet higher than the SO site is Cerro Chajnantor (5,612 m), the location selected for CCAT-prime [18] and TAO [19].

Because of these prominent projects, the Atacama Desert has been extensively characterized by various radiometers and far-IR tau-meters. The lower and higher sites are located within the same MERRA-2 grid and can be used to validate the elevation hypothesis: Setting the starting point of the integral in Eqn (1) to be the ground pressure of the site.

Extensive site surveys of the Chajnantor area have been carried out for ALMA [20]. Additionally, in a series of papers spanning more than a decade, Radford & Peterson reported distributions of *PWV* inferred by 350 µm continuum measurements carried out at Mauna Kea, Chajnantor, and South Pole [21,22,23].

The *PWV* distributions derived from MERRA-2 for SO and Cerro Chajnantor (Figure 5; Table 1) are comparable with these archival measurements. Although the *PWV* derived from the 350 µm tipper measurements for Cerro Chajnantor are noticeably lower than MERRA-2 (0.66 mm vs 0.75 mm) [22]. It is not clear whether the difference is simply due to yearly variations, or conversions from opacity to *PWV*. On the other hand, the APEX (5,104 m) 183 GHz water line data are in very good agreement with the MERRA-2 values for SO (median of 0.99 mm vs 1.1 mm).

The *PWV* scale height derived from MERRA-2 (1,475 m at the median) agrees well with [24], which reported an exponential scale height of 1,379 m at the median using two identical radiometers, but is somewhat larger than the value given in [23] of 1,080 m.

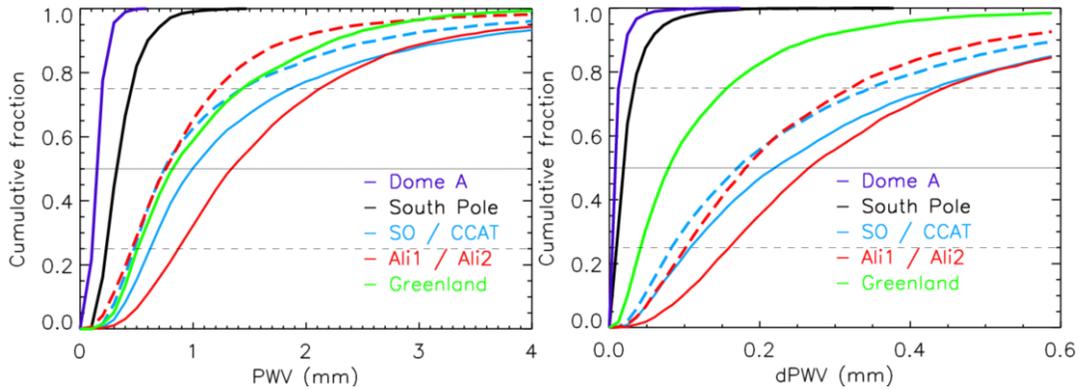

Figure 5. (*Left*) the cumulative distribution of precipitable water vapor for Dome A, South Pole, Chajnantor (SO/CCAT), Ali1/Ali2, and Summit Camp, Greenland. For Chajnantor and Ali, two sets of curves are given for lower (solid) and higher (dashed) sites. Greenland, Ali2 both provide similar observing conditions to Cerro Chajnantor. (*Right*) the cumulative distribution of *dPWV*, the fluctuation measure defined in equation (2). Dome A and South Pole are superb for both their low *PWV* quartiles and stability. Table 1 provides a numerical summary of Figure 5.

Table 1: Comparisons between mm/submm sites

| | elev. (m) | *PWV* (mm) | | | d*PWV* (mm) | | | *IWP* (kg.m^-2) | | | *LWP* (kg.m^-2) | | |
|---|---|---|---|---|---|---|---|---|---|---|---|---|---|
| | | 25% | **50%** | 75% | 25% | **50%** | 75% | 25% | **50%** | 75% | 25% | **50%** | 75% |
| Dome A | 4,093 | 0.105 | **0.141** | 0.191 | 3.65E-03 | **6.56E-03** | 1.21E-02 | 8.09E-05 | **2.30E-04** | 6.11E-04 | - | **-** | - |
| South Pole | 2,835 | 0.231 | **0.321** | 0.448 | 1.04E-02 | **1.77E-02** | 3.17E-02 | 1.96E-04 | **1.14E-03** | 3.97E-03 | - | **1.77E-05** | 4.54E-04 |
| Chajnantor (SO) | 5,190 | 0.618 | **0.993** | 1.871 | 1.07E-01 | **2.20E-01** | 4.32E-01 | - | **2.23E-05** | 1.69E-03 | - | **-** | 4.33E-05 |
| Cerro Chajnantor | 5,612 | 0.48 | **0.746** | 1.439 | 8.28E-02 | **1.71E-01** | 3.47E-01 | - | **2.15E-05** | 1.69E-03 | - | **-** | 3.22E-05 |
| Ali1 | 5,250 | 0.871 | **1.343** | 2.125 | 1.59E-01 | **2.66E-01** | 4.45E-01 | 7.62E-06 | **1.16E-03** | 1.14E-02 | - | **8.70E-04** | 8.91E-03 |
| Ali2 | 6,100 | 0.459 | **0.759** | 1.207 | 1.01E-01 | **1.81E-01** | 3.21E-01 | 1.70E-06 | **9.91E-04** | 1.08E-02 | - | **5.36E-04** | 7.56E-03 |
| Greenland | 3,216 | 0.509 | **0.817** | 1.436 | 4.14E-02 | **7.89E-02** | 1.56E-01 | 6.34E-04 | **2.54E-03** | 8.34E-03 | 5.18E-04 | **3.57E-03** | 1.15E-02 |

*South Pole* The geographical South Pole (*elev.* 2,835 m) is home to a series of successful CMB experiments since 1990's. US National Science Foundation provided excellent logistic support to the South Pole-based experiments. Currently, the BICEP3 telescope [25], Keck Array [26], and the 10 m South Pole Telescope (SPT) [27] are observing from the South Pole at mm-wave bands.

The observing conditions at the South Pole have been extensively characterized [21,23,28,29]. The dry, calm, and extremely cold conditions prevalent on the Antarctic Plateau allow it to surpass the higher Chajnantor in mm/submm atmospheric transparency. The *PWV* distributions derived from the MERRA-2 reanalysis are presented in Figure 5 and Table 1. The 2014–2016 8-month *PWV* quartiles of 0.231/ 0.321/0.448 mm affirm South Pole's tremendous advantage in uninterrupted mm-wave observations. These numbers are in excellent agreement with those reported elsewhere, *e.g.,* in reference [29], which estimated the best quartile and median *PWV* to be 0.23 mm and 0.32 mm.

Besides its low *PWV*, it has been established that the South Pole atmosphere is extremely stable with low sky temperature fluctuations due to lack of diurnal variations and its location away from the oceans [21]. It has been shown that the supreme stability of the sky brightness temperature at South Pole truly separates it from Chajnantor and other mid-latitude sites such as Mauna Kea [10,11, 30 , 31 ]. These direct observations of spatial fluctuations agree well with the simple, but uniform fluctuation measures proposed in this paper (Sections 2, 4.4).

4.2 *Basic Properties of the Candidate Sites*

*Dome A (Kunlun Station)* Dome Argus (4,093 m) is located at the highest point of the Antarctic Plateau, 1,260 m higher than the South Pole and further inland. Since early 2000, Australian, Chinese, and US groups identified Dome A as a potentially game-changing site for THz astronomy. This assessment was initially supported by weather data and some limited onsite measurements during the summer. The Chinese Kunlun Station was established at Dome A in 2009, four years after it was visited for the first time. Ever since, tremendous efforts were made to characterize the its THz transmittance in the winter, when temperatures and submillimeter opacities are lowest, and where the differences in conditions between sites are most significant. A Schottky Diode-based receiver was online for 204 days in 2008 to measure sky opacity at 661 GHz [29]. The inferred best quartile and median *PWV* are 0.1 mm and 0.14 mm, which are in good agreement with the MERRA-2 results presented in this paper (0.105 mm, 0.141 mm). Recently, 19 months of FTS data covering 0.75–15 THz are analyzed, revealing the THz windows promised by the low *PWV* [5].

In this paper, it is shown that the *PWV* quartiles at Dome A are more than a factor of two smaller than those of the South Pole, using data retrieved by the same satellite-borne instruments and using the same analysis method. The implication is that in addition to being a premier THz site, Dome A is superior to South Pole for long CMB integrations at high frequencies. For example, the 3$^{rd}$ quartile 220 GHz sky brightness temperature is 30% lower at Dome A than at the South Pole.

Furthermore, the quartile values for *dPWV* (fluctuations as defined in Section 2), *IWP* (cirrus clouds), and *LWP* (liquid clouds) are all much lower at Dome A than at South Pole (Figure 5 and Table 1). These will be discussed further in Sections 4.3 and 4.4.

*Summit Camp, Greenland*  Summit Camp is located at an elevation of 3,216 m in Greenland, 300–500 km inland. The current research areas at the station include atmospheric sciences, cosmic rays, and neutrino physics. CfA and ASIAA are collaborating on a major project that will install a 12 m submm antenna (GLT, Greenland Telescope [3]) for VLBI studies of the supermassive blackhole at the center of M87. Summit Camp is also a potential Northern Hemisphere site for the next stage of CMB programs. The extreme polar environment and lack of diurnal variations could in principle provide uninterrupted observing conditions similar to South Pole. Its latitude of 72.5°N makes it possible to form "cross-linked" measurements of the sky. However, continuous operation of CMB telescopes requires major improvements to the infrastructure and logistic support much beyond the level of GLT, which will only be online for 10% of the time.

The ASIAA team deployed a radiometer manufactured by Radiometer Physics GmbH (RPG) to monitor zenith 220 GHz opacity over the past few seasons. They reported that the inferred *PWV* is about 10% larger than those derived from their own MERRA-2 analyses. The measured *PWV* median of 0.79 mm for the 6 winter months over the last 3.5 years [8] is in good agreement with the results reported in this paper, which put the *PWV* quartiles at 0.509/0.817/1.436 mm. These quartile range is comparable to Cerro Chajnantor and Ali2 considered below (Figure 5; Table 1). However, MERRA-2 data indicate that liquid water is present at a surprisingly high level in Greenland. *LWP* and its fluctuations can significantly affect the efficiency of long integrations but are often overlooked in the literature (more discussions in Section 4.3 and 4.4).

*Ali, Tibet*  Ali is the site of a major Chinese initiative on CMB and primordial gravitational waves. Located on the dry side of the Himalayas at the latitude of 32°N, the site has access to a large portion of the sky, especially the Galactic northern hole, which is difficult to observe from Atacama because the field is available from the Southern Hemisphere often when the weather is less than ideal. Ali is easily accessed through daily commercial flights from Lhasa, Tibet's capital city. Ample electricity is available through the city grid, eliminating the need for generators on site and the logistic challenges of maintenance and transportation of diesel fuel. Furthermore, the site is about one hour from a major town of 20,000 people, Shiquanhe, where food and other supplies are available throughout the year. The town is still under fast development so construction services and equipment are easily obtained in the area.

The MERRA-2-derived *PWV* for Ali has been checked against radiosonde launches by the local weather station for the period between October–November 2016. The two data sets showed excellent agreements [ 32 ]. From MERRA-2 data, it is confirmed that Ali1 (5,250 m), the site planned for the first 90/150 GHz telescope, is a very decent mm-wave observing site, with *PWV* quartiles coming in at 0.871/1.343/2.125 mm between September and April. Although the median *PWV* is somewhat larger than that of SO (0.993 mm), the 85% cumulative *PWV* value is nearly the same as that of the SO. Since 4 worse summer months are already excluded, this is perhaps a more representative comparison for high-duty cycle observations such as CMB.

In Figure 5 and Table 1, the properties for Ali2, a proposed site at 6,100m within the same MERRA-2 grid are also presented. The inferred *PWV* quartiles (0.459/0.759/1.207 mm) are very similar to Cerro Chajnantor and Summit Camp, Greenland. The derived *PWV* scale height for the Ali area is 1,490 m at the median. The level of *LWP* is found to be higher in Ali than in Atacama, though not as high as Summit Camp. Ali2 observatory, when established, will be a premier mm/submm site at the mid-latitude on the Northern Hemisphere.

## 4.3 Impact of Ice and Liquid Clouds

As discussed earlier, MERRA-2 retrieves *PWV*, *LWP*, and *IWP* separately using measurements over a wide frequency range. In Figure 4, optical depth due to various components of the atmosphere is shown for representative values of *PWV* (0.64mm) and *LWP* (0.01 kg/m$^2$) in winter at the Summit Camp. At 90 and 150 GHz, the contribution from liquid clouds is comparable to the water vapor. Liquid affects a much smaller fraction of the data at the South Pole because the *LWP* median is ∼ 200× smaller. (Figure 6a; Table 1).

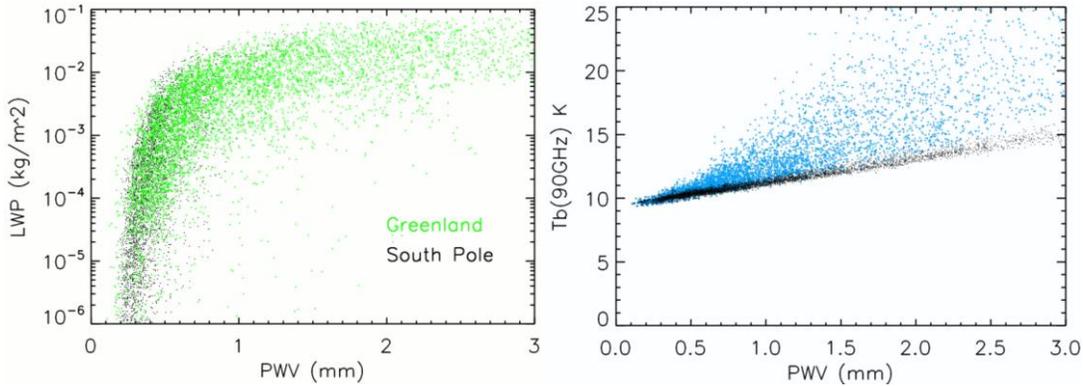

Figure 6. (*Left*) The liquid water path (*LWP*) plotted against precipitable water vapor (*PWV*) for South Pole and Greenland. Liquid water starts to become an important factor when *PWV* >∼ 0.5 mm. (*Right*) The zenith brightness temperature at 90 GHz vs *PWV* in Greenland. When *LWP* is ignored, $T_b$ is roughly linearly proportional to the *PWV* (black dots). When *PWV* is larger, *LWP* adds important contributions to $T_b$ (cyan dots). Also see Figure 4.

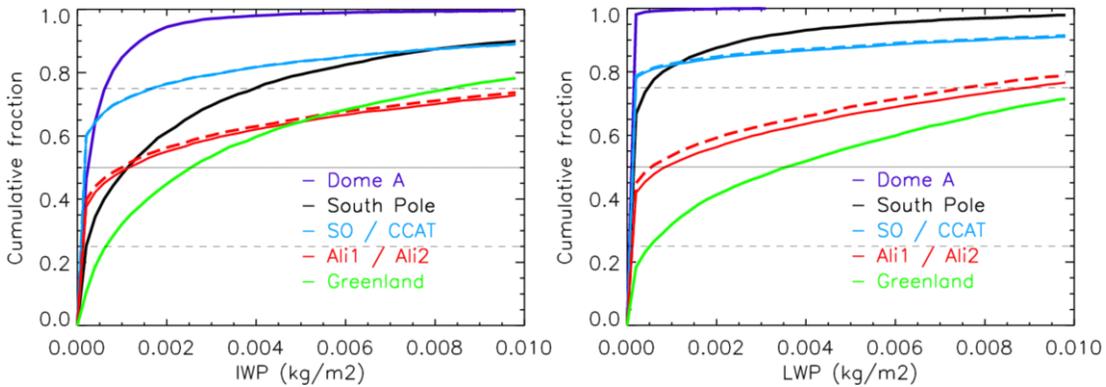

Figure 7. The cumulative distribution of ice water path, *IWP* (*Left*) and liquid water path, *LWP* (*Right*) for the five sites considered in this study. As discussed in the text, liquid water clouds will have an important impact on observations.

In Figure 6b, *am* values of zenith brightness temperature at 90 GHz are plotted against *PWV* with (cyan) and without (black) the effects of liquid water. When *PWV* is low, sky brightness temperature is proportional to *PWV*. However, when *PWV* creeps above ~0.5 mm, supercooled liquid droplets appear and become a hidden variable that add significantly to $T_b$ – and increase, or even *dominate* its fluctuations (Section 4.4).

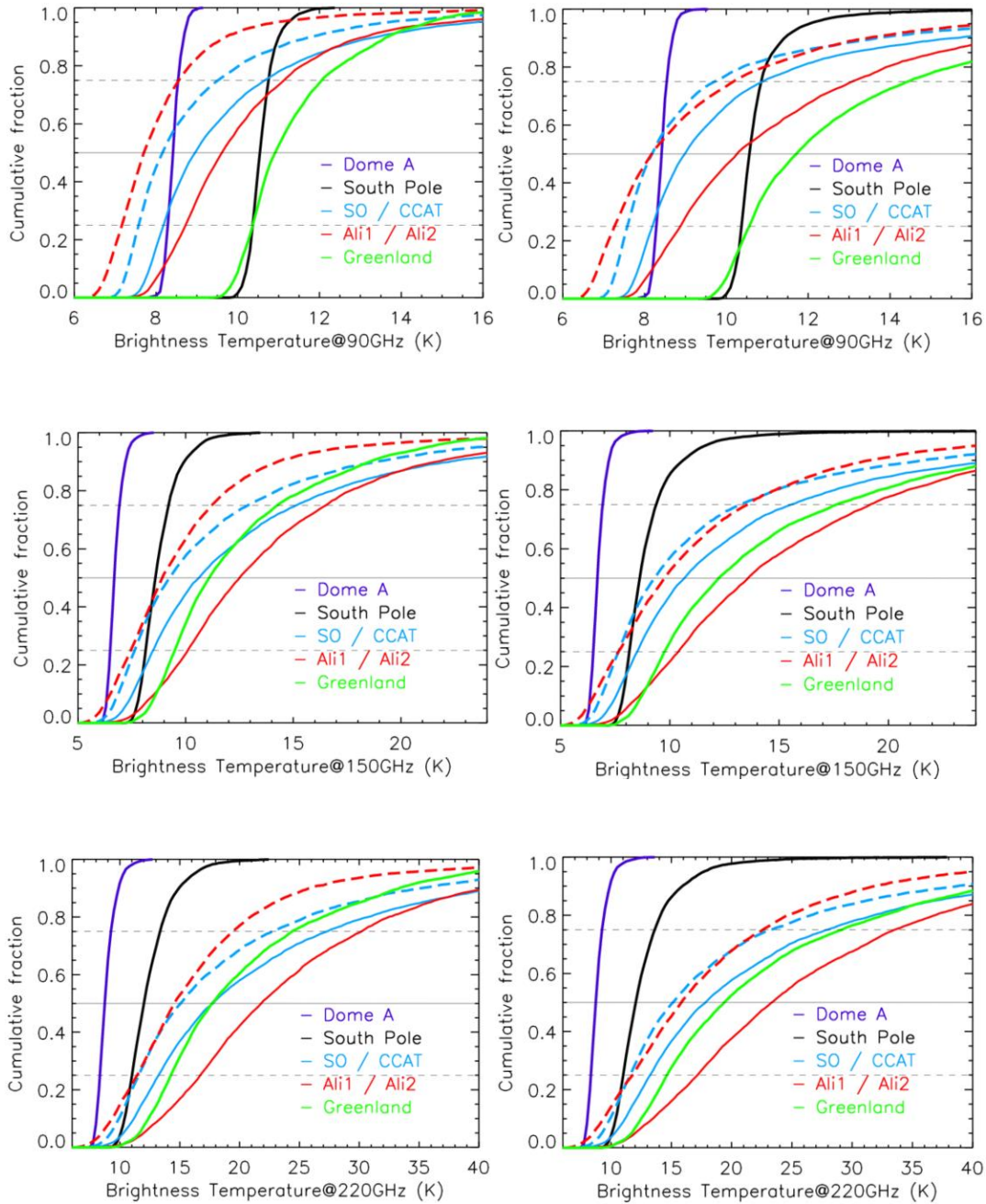

Figure 8. Zenith brightness temperature (Planck) at 90, 150, 220 GHz calculated by *am* radiative transfer code for the five sites considered in this paper. The calculations are repeated with (*Left* column) and without (*Right* column) including effects of liquid and ice water clouds. Table 2 is the numerical summary of this Figure.

Table 2. Zenith Planck Brightness Temperature $T_b$

|  |  | 90GHz | | | 150GHz | | | 220GHz | | |
|---|---|---|---|---|---|---|---|---|---|---|
|  |  | 25% | 50% | 75% | 25% | 50% | 75% | 25% | 50% | 75% |
| Dome A | vapor only | 8.29 | 8.42 | 8.53 | 6.50 | 6.69 | 6.93 | 8.33 | 8.73 | 9.25 |
|  | vp+ice+liq. | 8.29 | 8.42 | 8.53 | 6.50 | 6.69 | 6.93 | 8.34 | 8.73 | 9.25 |
| South Pole | vapor only | 10.37 | 10.54 | 10.76 | 8.14 | 8.59 | 9.20 | 11.01 | 12.01 | 13.36 |
|  | vp+ice+liq. | 10.37 | 10.56 | 10.84 | 8.14 | 8.62 | 9.37 | 11.01 | 12.05 | 13.61 |
| Simons Obs. | vapor only | 8.19 | 8.96 | 10.67 | 8.48 | 10.62 | 15.15 | 13.20 | 17.77 | 27.29 |
|  | vp+ice+liq. | 8.19 | 8.98 | 10.90 | 8.50 | 10.67 | 15.57 | 13.22 | 17.85 | 27.97 |
| Cerro Chaj. | vapor only | 7.57 | 8.16 | 9.52 | 7.63 | 9.27 | 12.89 | 11.57 | 15.08 | 22.75 |
|  | vp+ice+liq. | 7.58 | 8.18 | 9.71 | 7.65 | 9.30 | 13.25 | 11.60 | 15.15 | 23.27 |
| Ali1 | vapor only | 8.72 | 9.61 | 11.15 | 10.09 | 12.50 | 16.53 | 16.70 | 21.81 | 30.19 |
|  | vp+ice+liq. | 8.88 | 10.26 | 13.07 | 10.40 | 13.57 | 19.16 | 17.18 | 23.36 | 33.50 |
| Ali2 | vapor only | 7.16 | 7.70 | 8.56 | 7.47 | 8.91 | 11.22 | 11.46 | 14.55 | 19.43 |
|  | vp+ice+liq. | 7.27 | 8.18 | 10.17 | 7.65 | 9.75 | 13.56 | 11.78 | 15.83 | 22.57 |
| Greenland | vapor only | 10.36 | 10.92 | 12.05 | 9.49 | 11.10 | 14.25 | 14.25 | 17.70 | 24.40 |
|  | vp+ice+liq. | 10.56 | 11.72 | 14.44 | 9.75 | 12.23 | 17.61 | 14.62 | 19.38 | 29.14 |

Table 3. *PWV* and *LWP* fluctuations

|  | $\Delta PWV$ (mm) | | | $\Delta LWP$ (kg.m^-2) | | |
|---|---|---|---|---|---|---|
|  | 25% | 50% | 75% | 25% | 50% | 75% |
| South Pole | 1.45E-03 | 2.76E-03 | 5.29E-03 | - | 1.00E-05 | 9.13E-05 |
| Greenland | 6.06E-03 | 1.30E-02 | 2.78E-02 | 1.33E-04 | 5.47E-04 | 1.57E-03 |

Table 4. Zenith (Rayleigh-Jeans) Temperature Fluctuations

|  |  |  | 90GHz | | | 150GHz | | | 220GHz | | |
|---|---|---|---|---|---|---|---|---|---|---|---|
|  |  |  | 25% | 50% | 75% | 25% | 50% | 75% | 25% | 50% | 75% |
| Dome A | $dT_{rj}$ (K) |  | 5.6E-03 | 1.0E-02 | 1.9E-02 | 1.9E-02 | 3.5E-02 | 6.4E-02 | 4.2E-02 | 7.6E-02 | 0.14 |
| South Pole | $dT_{rj}$ (K) |  | 1.6E-02 | 2.7E-02 | 4.9E-02 | 5.5E-02 | 9.3E-02 | 0.17 | 0.12 | 0.20 | 0.37 |
|  | $\Delta T_{rj}$ (K) | vapor only | 2.24E-03 | 4.27E-03 | 8.18E-03 | 7.65E-03 | 1.46E-02 | 2.79E-02 | 1.67E-02 | 3.19E-02 | 6.11E-02 |
|  |  | v+liq. lnr | 3.10E-03 | 7.08E-03 | 2.07E-02 | 9.55E-03 | 2.04E-02 | 5.08E-02 | 1.99E-02 | 4.21E-02 | 9.61E-02 |
|  |  | v+liq. quad | 2.77E-03 | 6.18E-03 | 1.65E-02 | 8.73E-03 | 1.80E-02 | 4.13E-02 | 1.84E-02 | 3.65E-02 | 8.04E-02 |
| Simons Obs. | $dT_{rj}$ (K) |  | 0.17 | 0.34 | 0.67 | 0.55 | 1.14 | 2.24 | 1.20 | 2.46 | 4.83 |
| Cerro Chaj. | $dT_{rj}$ (K) |  | 0.13 | 0.26 | 0.54 | 0.43 | 0.89 | 1.80 | 0.93 | 1.91 | 3.89 |
| Ali1 | $dT_{rj}$ (K) |  | 0.25 | 0.41 | 0.69 | 0.82 | 1.38 | 2.31 | 1.78 | 2.98 | 4.98 |
| Ali2 | $dT_{rj}$ (K) |  | 0.16 | 0.28 | 0.50 | 0.52 | 0.94 | 1.66 | 1.13 | 2.03 | 3.59 |
| Greenland | $dT_{rj}$ (K) |  | 6.39E-02 | 0.12 | 0.24 | 0.21 | 0.41 | 0.81 | 0.46 | 0.88 | 1.75 |
|  | $\Delta T_{rj}$ (K) | vapor only | 9.35E-03 | 2.01E-02 | 4.28E-02 | 3.14E-02 | 6.75E-02 | 0.14 | 6.79E-02 | 0.15 | 0.31 |
|  |  | v+liq. lnr | 3.40E-02 | 0.11 | 0.29 | 8.00E-02 | 0.21 | 0.50 | 0.14 | 0.36 | 0.81 |
|  |  | v+liq. quad | 3.01E-02 | 9.41E-02 | 0.25 | 6.49E-02 | 0.17 | 0.40 | 0.11 | 0.27 | 0.64 |

Figure 6b clearly demonstrates that simultaneous measurements of *PWV* and *LWP* are necessary for a proper evaluation of the site, even though *LWP* is often unjustifiably omitted. Cumulative distributions of *IWP* and *LWP* are presented in Figure 7 and Table 1 for different sites. One can notice that in these data, *IWP* and *LWP* often come in at a similar level. Since ice crystals scatter less and absorb less than liquid, they do not significantly impact mm-wave observations and are listed only for completeness.

In Figure 8 and Table 2, zenith Planck brightness temperature $T_b$ calculated by *am* is shown for three key CMB bands 90, 150, and 220 GHz, with and without effects of *IWP* and *LWP*. Liquid water is nearly undetectable at Dome A, and only plays a minor role at the South Pole and Atacama. On the other hand, the presence of liquid water noticeably increases the sky temperature in Greenland, especially at the two lower frequency bands important for the CMB. The level of *LWP* in Ali lies between Atacama and Greenland.

*4.4 Brightness Temperature Fluctuations*

Prior work on quantifying brightness temperature fluctuations (a.k.a. the "sky noise") for various sites often adopts specific models of the atmospheric fluctuations. Estimations for the parameters in the model, such as fluctuation amplitudes or power law index are then derived based on the observed spatial or temporal fluctuations [10,11,30,31]. The advantage of that approach is that these parameters are directly related to quantitative models. However, different papers often chose different models, using datasets obtained under different conditions. Such variations severely complicate direct comparisons of different sites, which is a main purpose of these studies. The "$d$" measures proposed in this paper based on spatial variations of specific humidity (equation 2) should provide simple and robust comparisons between sites. Normalized by the $dPWV$ median of South Pole, the $dPWV$ medians for Dome A, Greenland, Cerro Chajnantor, Simons Obs., Ali2, Ali1 are 0.37, 4.5, 9.7, 12.4, 10.2, 15.0, respectively (Table 1). These ratios correspond roughly to Rayleigh-Jeans brightness temperature fluctuations ($dT_{rj}$) due to variations in water vapor (Table 4). The squares of these quantities represent the fluctuation power against which $r$ (primordial tensor-to-scalar ratio) measurements are made.

As described in Section 2, the alternative $\Delta$ quantities have been derived from 2D data with 1-hr intervals. Table 3 shows that the median $\Delta PWV$ for Greenland is $4.7\times$ larger than for the South Pole. This is consistent with the factor of 4.5 quoted above for $dPWV$, giving us confidence that both measures are probing the same underlying Kolmogorov-Taylor spectra. For Summit Camp, Greenland, *LWP* and its fluctuations play an important role. The $\Delta$ measures of *PWV* and *LWP* provide an easy way to quantify their relative contributions to brightness temperature fluctuations. Feeding these into the *am* radiative transfer code, one can obtain approximate brightness temperature $T_{rj}$ fluctuations, as listed in Table 4.

How $\Delta PWV$ and $\Delta LWP$ add together depends on the correlation between these quantities on relevant time scales. In Table 4, both linear (100% correlation) and quadrature sums (uncorrelated) are presented. It is noted that for Greenland, the brightness temperature fluctuations due to vapor and liquid are much larger than those from the vapor alone, even if quadrature sums are used (At 90 GHz and 150 GHz, $\Delta T_{rj}$ is at least $\sim 10\times$ larger at Summit Camp than at the South Pole).

## 5. DISCUSSION

Using MERRA-2 reanalysis of atmospheric data, several remote sites are comprehensively evaluated for mm-wave observations, with an emphasis on long integrations with high duty cycles. The numerical results have been summarized in Tables 1–4. The properties evaluated include the standard *PWV* distribution, its variations (spatial and temporal) on time/angular scales relevant to future CMB observations, and the statistics of ice and liquid clouds. This work expands the scope of prior site characterization, and eliminates equipment and methodology-related uncertainties.

Dome A is confirmed to be superior to South Pole in all measures, including *PWV* quartiles, sky brightness, and its fluctuations. It can be a game-changer for Galactic foreground measurements at frequencies higher than 280 GHz, which is challenging even for South Pole.

Ali1 at 5,250 m is an excellent site for the planned first-stage observations at 90/150 GHz. The *PWV* median and best quartile are somewhat higher than those of Atacama at the same elevation so it is motivated to search for a >6,000 m site for observations at 220 GHz and higher frequencies. Fortunately, there are several easily accessible peaks above 6,000 m between the airport and the current Ali1 site. Such high site would have *PWV* quartiles and stability nearly identical to those at Cerro Chajnantor (5,612 m). Plans are being made to place automatic weather stations on these candidate peaks.

Based on the distributions of *PWV*, Summit Camp, Greenland provides observing conditions similar to Cerro Chajnantor and Ali2. Perhaps the biggest surprise from MERRA-2 reanalysis is the relative high level of liquid water clouds detected in Greenland. Not only does the *LWP* contribute significantly to 90/150 GHz opacity, the sky temperature fluctuations caused by variations in *LWP* often dominate over those from *PWV* variations. This is analogous to how massive galaxies of clusters present a highly amplified version of the matter distribution of the universe. Based on MERRA-2 data, the sum effects of vapor and liquid generate brightness temperature fluctuations at the level similar to the mid-latitude sites (10× that of the South Pole). Since radiometer observing only at around the 183 GHz water line would miss *LWP* and its fluctuations altogether, this result calls for simultaneous retrieval of *PWV* and *LWP* in future site surveys. It is also worth reviewing how *PWV, IWP,* and *LWP* are retrieved in MERRA-2 and how these quantities are correlated.

6. ACKNOWLEDGEMENT

The author thanks Keiichi Asada, Nien-Ying Chou, Ken Ganga, Paul Ho, Adrian Lee, Yongping Li, Yang Liu, Shengcai Shi, Keith Thompson, Chun-Hao To, Suzanne Staggs, Wei-Hsin Sun, and Xinmin Zhang for discussions and useful suggestions to an early manuscript. The author also thanks the support from SCPKU (Stanford Center at Peking University), where a large part of this paper was written. Some of the analyses are done using Goddard IDL library.